\documentclass[twocolumn,english,prb,twocolumn,nofootinbib]{revtex4-1}
\usepackage[T1]{fontenc}
\usepackage[latin9]{inputenc}
\usepackage{babel}
\usepackage{amsmath}
\usepackage{amssymb}
\usepackage{graphicx}
\usepackage{esint}
\usepackage[unicode=true,pdfusetitle,
 bookmarks=false,
 breaklinks=false,pdfborder={0 0 0},backref=false,colorlinks=false]
 {hyperref}
\usepackage{breakurl}

\makeatletter
\usepackage{babel}
\usepackage{babel}
\usepackage{dcolumn}

\makeatother

\begin{document}

\title[First-order kinetics bottleneck during photoinduced ultrafast insulator-metal
transition ...]{First-order kinetics bottleneck during photoinduced ultrafast insulator-metal
transition in 3D orbitally-driven Peierls insulator CuIr$_{2}$S$_{4}$ }

\author{M. Naseska$^{1}$, P. Sutar$^{1}$, Y. Vaskivskyi$^{1}$, I. Vaskivskyi$^{1}$,
D. Vengust$^{1}$, D. Svetin$^{1}$, V. V. Kabanov$^{1}$, D. Mihailovic$^{1,2}$,
T. Mertelj$^{1,2}$}

\address{$^{1}$Complex Matter Department, Jozef Stefan Institute, Jamova
39, 1000 Ljubljana, Slovenia}

\address{$^{2}$Center of Excellence on Nanoscience and Nanotechnology Nanocenter
(CENN Nanocenter), Jamova 39, 1000 Ljubljana, Slovenia}

\email{* tomaz.mertelj@ijs.si}

\begin{abstract}
The spinel-structure CuIr$_{2}$S$_{4}$ compound displays a rather
unusual orbitally-driven three-dimensional Peierls-like insulator-metal
transition. The low-$T$ symmetry-broken insulating state is especially
interesting due to the existence of a metastable irradiation-induced
disordered weakly conducting state. Here we study intense femtosecond
optical pulse irradiation effects by means of the all-optical ultrafast
multi-pulse time-resolved spectroscopy. We show that the structural
coherence of the low-$T$ broken symmetry state is strongly suppressed
on a sub-picosecond timescale above a threshold excitation fluence
resulting in a structurally inhomogeneous transient state which persists
for several-tens of picoseconds before reverting to the low-$T$ disordered
weakly conducting state. The electronic order shows a transient gap
filling at a significantly lower fluence threshold. The data suggest
that the photoinduced-transition dynamics to the high-$T$ metallic
phase is governed by first-order-transition nucleation kinetics that
prevents the complete ultrafast structural transition even when the
absorbed energy significantly exceeds the equilibrium enthalpy difference
to the high-$T$ metallic phase. In contrast, the dynamically-decoupled
electronic order is transiently suppressed on a sub-picosecond timescale
rather independently due to a photoinduced Mott transition. 
\end{abstract}
\maketitle
\noindent \textit{Keywords\/}: {ultrafast metal-insulator transition,
ultrafast optical spectroscopy, transient reflectivity, coherent phonons,
CuIr$_{2}$S$_{4}$, spinel}

\section{Introduction}

Kinetics of first order phase transitions are important both from
the point of view of applications as well as fundamental science.
For example, the control of kinetics during the transformation of
various forms of steel \citep{papon2002physics} is used to tailor
its microstructural and mechanical properties. An example of interesting
fundamental physics related to first order phase transition kinetics
is inflation of the early universe, where a supercooled metastable
system might been transformed via droplet nucleation into the Higgs-field
broken-symmetry state \citep{linde1982new}.

In solids, ultrafast first-order insulator-metal (IM) phase transitions
could be instrumental for ultrafast sensor and nonvolatile memory
applications. Their ultrafast kinetics therefore attracted a great
deal of attention \citep{beckerBuckman1994femtosecond,fiebigMiyano2000ultrafast,cavalleriToth2001femtosecond,perfettiLoukakos2006time,BaumYang2007,wallWegkamp2012ultrafast,deJongKukreja2013,fukazawaTanaka2013time,stojchevskaVaskivskyi2014ultrafast,morrisonChatelain2014photoinduced,wegkampHerzog2014instantaneous,abreuWang2015dynamic,OcallahanJones2015,hauptEichberger2016ultrafast,zhangTan2016cooperative,lantzMansart2017ultrafast,jager2017tracking,laulheHuber2017ultrafast,singerRamirez2018nonequilibrium,ligges2018ultrafast,ronchiHomm2019early,WallYang2019,OttoCotret2019,vidasSchick2020}
with strong focus on VO$_{2}$ \citep{beckerBuckman1994femtosecond,cavalleriToth2001femtosecond,BaumYang2007,wallWegkamp2012ultrafast,wegkampHerzog2014instantaneous,morrisonChatelain2014photoinduced,OcallahanJones2015,jager2017tracking,WallYang2019,OttoCotret2019,vidasSchick2020},
V$_{2}$O$_{3}$ \citep{abreuWang2015dynamic,lantzMansart2017ultrafast,singerRamirez2018nonequilibrium,ronchiHomm2019early}
and 1\emph{T}-TaS$_{2}$ \citep{perfettiLoukakos2006time,stojchevskaVaskivskyi2014ultrafast,hauptEichberger2016ultrafast,laulheHuber2017ultrafast,ligges2018ultrafast}.
There are however still open questions how the inherent first-order
meta-stability manifests itself when phases with concurrent electronic
and lattice orders are driven across the first-order phase boundary
on ultrafast time scales.

In VO$_{2}$, for example, the change of the low-$T$ electronic-order-induced
monoclinic lattice potential is believed to be rather abrupt \citep{BaumYang2007,wallWegkamp2012ultrafast,jager2017tracking,WallYang2019}
with a sub-picosecond V-V dimerization suppression suggesting that
the high-$T$ metallic rutile phase symmetry is restored on a $\sim100$-fs
timescale. This is contrasted by recent observation of an additional
transient metastable monoclinic metallic phase in a significant fraction
of the polycrystaline sample grains \citep{OttoCotret2019}. While
the presence of the metastable monoclinic metallic phase is still
rather controversial \citep{vidasSchick2020} the electronic and lattice
orders might be transiently decoupled with the emergence of the metallic
rutile lattice structure somehow inhibited.

In V$_{2}$O$_{3}$ the inherent first-order meta stability plays
a determining role with the nucleation and growth kinetics of the
metallic nanodroplets governing the dynamics across the first-order
IM phase transition \citep{abreuWang2015dynamic,ronchiHomm2019early}.
The importance of the nucleation and growth kinetics was noted also
in a first-order ultrafast charge-density-wave phase transformation
in $1T$-TaS$_{2}$ \citep{hauptEichberger2016ultrafast,laulheHuber2017ultrafast}.

\begin{figure}
\includegraphics[width=1\columnwidth]{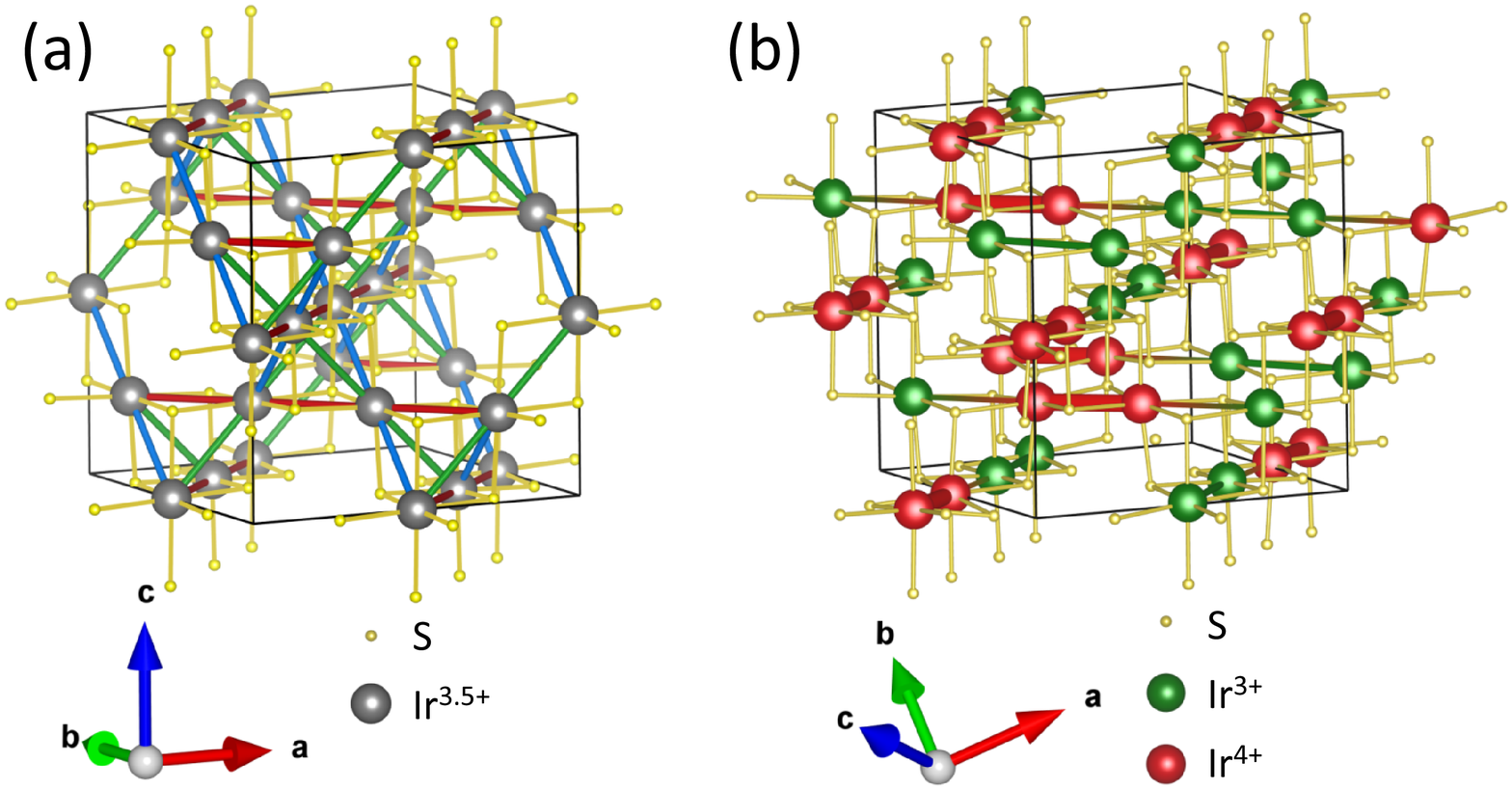}

\includegraphics[width=1\columnwidth]{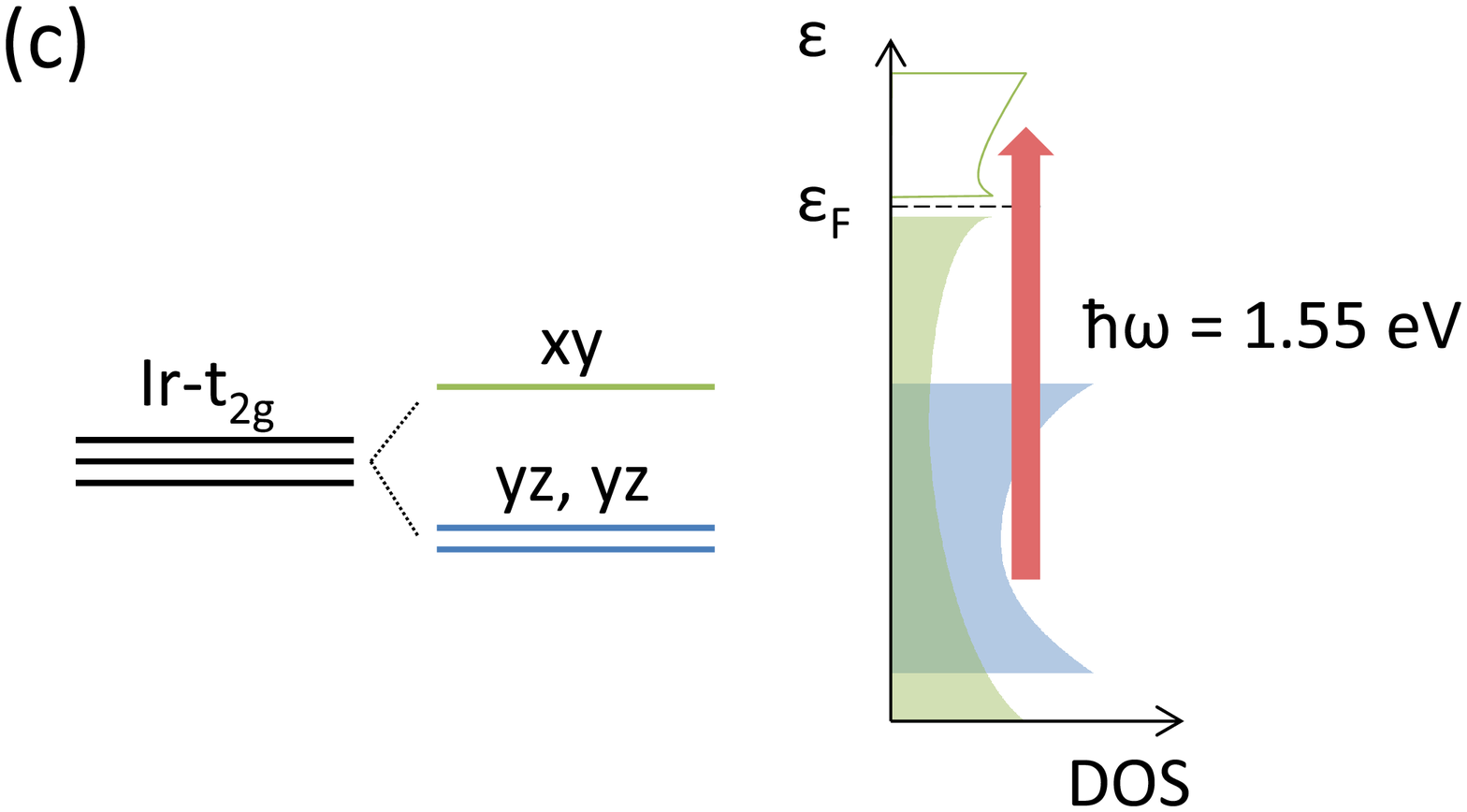}\caption{CuIr$_{2}$S$_{4}$ structure. (a) Schematics \citep{momma2011vesta}
of the CuIr$_{2}$S$_{4}$ structure in the high-$T$ metallic phase.
The different sets of Ir-ion rows alternating along the $\left\langle 110\right\rangle $
or $\left\langle -110\right\rangle $ cubic directions are indicated
by the differently colored bonds. (b) In the low-$T$ insulating phase
one set of rows undergoes a Peierls distortion with the Ir$^{4+}$
dimers (red dumbbells) interleaved with Ir$^{3+}$-ion pairs (green).
The compasses indicate the basis orientations while the black lines
indicate the high-$T$ cubic unit cell. The Cu ions are omitted for
clarity. (c) A schematic diagram of the Ir-t$_{2\mathrm{g}}$-orbitals
derived quasi-1D electronic bands density of states in the low-$T$
phase \citep{khomskiiMizokawa2005}. The shaded regions correspond
to the occupied states and the dominant optical transitions at the
experimental photon energy\citep{sarkarRaychaudhury2009density} are
indicated by the red arrow.\label{fig:struct}}
\end{figure}

An interesting opportunity for studying the ultrafast first-order
IM phase transition \citep{nagataHagino1994} kinetics is offered
in the spinel-structure CuIr$_{2}$S$_{4}$ compound that shows a
rather unusual orbitally-driven \citep{khomskiiMizokawa2005} three-dimensional
Peierls-like IM transition. Upon cooling, the metallic cubic (MC)
phase (Figure \ref{fig:struct} (a)) undergoes a first order transition
to a triclinic \citep{ishibashiSakai2001x,radaelliHoribe2002} insulating
phase (Figure \ref{fig:struct} (b)) at $T_{\mathrm{IM}}\sim$233
K. The triclinic insulating phase is characterized by Ir$^{3+}$:Ir$^{4+}$
charge disproportionation accompanied with Peierls-like dimerization
of Ir$^{4+}$-ion pairs along a set of cubic $\left\langle 110\right\rangle $
and $\left\langle -110\right\rangle $ Ir-ion rows alternating along
the corresponding $\left\langle 001\right\rangle $ direction.

The transition can be understood in terms of quasi-1D Ir-t$_{2g}$-orbitals
derived hybridized bands that are split due to the band Jahn-Teller
effect into two narrower fully occupied $xz$- and $yz$-derived bands
and a broader 3/4 occupied $xy$-derived band that simultaneously
becomes gaped due to the Peierls tetramerization (see Fig \ref{fig:struct}
(c)) \citep{khomskiiMizokawa2005}. The Ir-$5d$ orbitals appear broad
enough that the correlation effects do not play a major role in the
transition \citep{sasakiArai2004band,khomskiiMizokawa2005,sarkarRaychaudhury2009density}
contrary to V$_{2}$O$_{3}$ and VO$_{2}$.

The CuIr$_{2}$S$_{4}$ is interesting also due to the low temperature
X-ray- and visible-light-induced metastable disordered weakly conducting
(DWC) phase \citep{ishibashiKoo2002,furubayashiSuzuki2003,takuboHirata2005,kiryukhinHoribe2006,takuboMizokawa2008,bozinMasadeh2011}.
The presence of short-range incommensurate structural correlations
observed \citep{kiryukhinHoribe2006} in the DWC phase suggests presence
of \emph{competing instabilities} that appear after suppression of
the insulating phase. The formation of the DWC phase was studied mostly
with weak continuous excitations. Competing instabilities, however,
might lead to formation of even more conducting metastable state under
strongly nonequilibrium conditions as in the case of 1$T$-TaS$_{2}$,
where a strong ultrafast excitation leads to a metastable hidden (H)
metallic phase \citep{stojchevskaVaskivskyi2014ultrafast}. In order
to search for a possible hidden metallic state, different from the
DWC phase, and to study kinetics during ultrafast first-order IM phase
transition in CuIr$_{2}$S$_{4}$ we therefore performed a systematic
femtosecond multi-pulse all-optical investigation of the single-crystal
transient reflectivity relaxation dynamics in CuIr$_{2}$S$_{4}$
as a function of excitation fluence.

The low-$T$ quasi-equilibrium transient reflectivity in CuIr$_{2}$S$_{4}$
is dominated by the low-frequency broken-symmetry-induced coherent-phonons
response that discontinuously vanishes across the equilibrium IM transition.
\citep{NaseskaSutar2020} The coherent response can therefore be used
as a time-resolved probe for the broken-lattice-symmetry dynamics
in the strongly excited low-$T$ phases. \citep{yusupovMertelj2010,wallWegkamp2012ultrafast}
This can be done by means of our advanced multi-pulse method \citep{yusupovMertelj2010},
where we use a strong driving laser pulse to \emph{initiate} a photoinduced
transition and \emph{probe} the subsequent evolution by means of a
weaker pump- (P) probe (Pr) pulses sequence.

\section{Methods}

\subsection*{Sample growth and characterization}

Single crystals of CuIr$_{2}$S$_{4}$ were grown from Bi solution
and characterized as described in Ref. \citep{NaseskaSutar2020}.
Here we present ultrafast optical data from two cleaved crystals designated
S1 and S2. The orientation of the S1 cleaved surface was determined
from electron back scatter diffraction Kikuchi patterns to be close
to the $\left\langle 221\right\rangle $ plane while the orientation
of the S2 cleaved surface was inferred from the Raman selection rules
to be close to the $\left\langle 001\right\rangle $ plane. \citep{NaseskaSutar2020}

\subsection*{Multi-pulse transient reflectivity measurements}

\begin{figure}
\includegraphics[width=1\columnwidth]{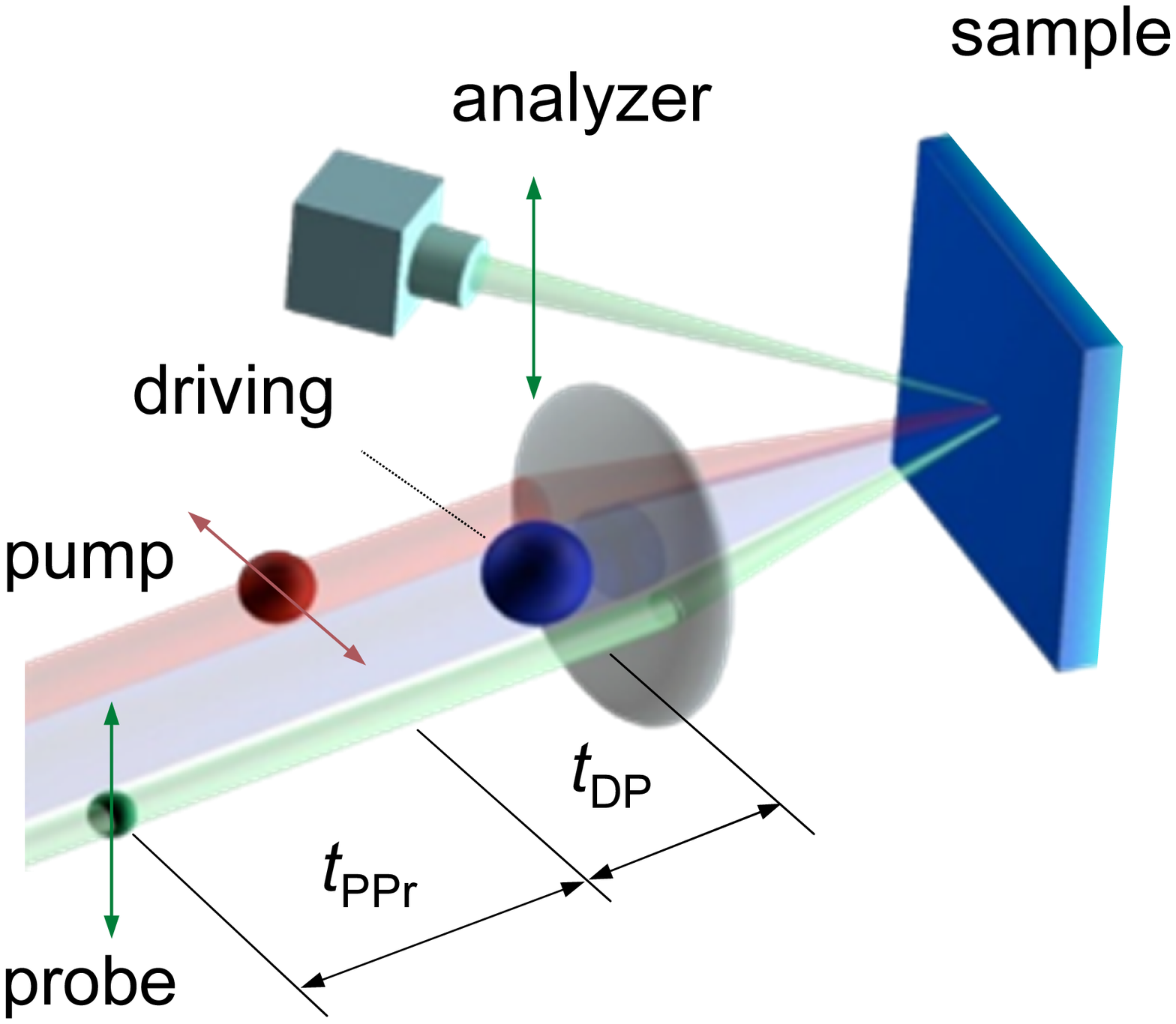}

\caption{Schematics of the multi-pulse experiment with notation of the delays
adapted from Ref. \citep{madanBaranov2017dynamics}.}
\label{fig:3p} 
\end{figure}

The multi-pulse transient reflectivity measurements. \citep{yusupovMertelj2010,naseskaPogrebna2018}
were performed using 50-fs linearly polarized laser pulses at 800
nm wavelength and the $200-250$ kHz repetition rate. In addition
to the pump (P) and probe (Pr) pulses at $\hbar\omega=1.55$~eV we
used another intense driving (D) pulse (also at $\hbar\omega=1.55$~eV)
with a variable delay with respect to the pump (P) pulse (see Figure
\ref{fig:3p}).

The multi-pulse transient reflectivity $\Delta R_{3}/R$ was measured
by monitoring the intensity of the Pr beam. The direct contribution
of the unchopped D beam to the total transient reflectivity, $\Delta R$,
was rejected by means of a lock-in synchronized to the chopper that
modulated the intensity of the P beam only. Due to the chopping scheme
the measured quantity in the multi-pulse experiments is the difference
between the transient reflectivity in the presence of P and D pulses,
$\Delta R_{\mathrm{DP}}(t_{\mathrm{Pr}},t_{\mathrm{P}},t_{\mathrm{D}})$,
and the transient reflectivity in the presence of the D pulse only,
$\Delta R_{\mathrm{D}}(t_{\mathrm{Pr}},t_{\mathrm{D}})$: 
\begin{equation}
\Delta R_{3}(t_{\mathrm{Pr}},t_{\mathrm{P}},t_{\mathrm{D}})=\Delta R_{\mathrm{DP}}(t_{\mathrm{Pr}},t_{\mathrm{P}},t_{\mathrm{D}})-\Delta R_{\mathrm{D}}(t_{\mathrm{Pr}},t_{\mathrm{D}}),\label{eq:DR3}
\end{equation}
where $t_{\mathrm{Pr}}$, $t_{\mathrm{P}}$ and $t_{\mathrm{D}}$
correspond to the Pr, P and D pulse arrival times, respectively. In
the limit of vanishing D pulse fluence $\Delta R_{3}/R$ reduces to
the standard two-pulse transient reflectivity $\Delta R/R$.

The P/D and Pr beam diameters were in the ranges of 40-70 and 18-30
$\mu$m, respectively. The probe fluence was $\sim5$0~$\mu$J/cm$^{2}$.
For the multi-pulse measurements the fluence of the P pulse, $\mathcal{F_{\mathrm{P}}}\lesssim100$~$\mu$J/cm$^{2}$,
was kept in the linear response region. The polarizations of the P
and D beams were perpendicular to the probe beam polarization with
a random orientation with respect to the crystal axes.

\section{Results}

The DWC phase emergence depends on the absorbed dose \citep{furubayashiSuzuki2003,koshimizuTsukahara2009formation},
but the threshold dose was determined only for the case of high-energy
ion irradiation\citep{koshimizuTsukahara2009formation}. Since the
exact optical DWC-phase-creation conditions are not known we measured
also the low-$T$ DC photoconductivity (see Supplemental Material
(SM) \citep{naseskaSutar2020suppl}) and found, as suggested previously,\citep{NaseskaSutar2020}
that even at the lowest feasible excitation fluences, $F\sim$100
$\mu$J/cm$^{2}$, the threshold dose is exceeded on a timescale faster
than a single transient reflectivity scan acquisition time of $\sim$100
s. The pristine insulating triclinic phase is therefore inaccessible
in the pump-probe experiments at low $T$ and the previously reported\citep{NaseskaSutar2020}
equilibrium low-$T$ transient reflectivity refers to the DWC phase.
Despite the absence of a long range Ir$^{4+}$-dimer order in the
DWC phase\citep{kiryukhinHoribe2006} the previously characterized
coherent oscillations (CO) due to the displacive excitation of coherent
phonons\citep{zeiger1992theory,NaseskaSutar2020,naseskaSutar2020suppl}
(DECP) show well defined peaks with the low-$T$ dephasing times exceeding
$\sim100$ ps.

\begin{figure}
\includegraphics[clip,width=1\columnwidth]{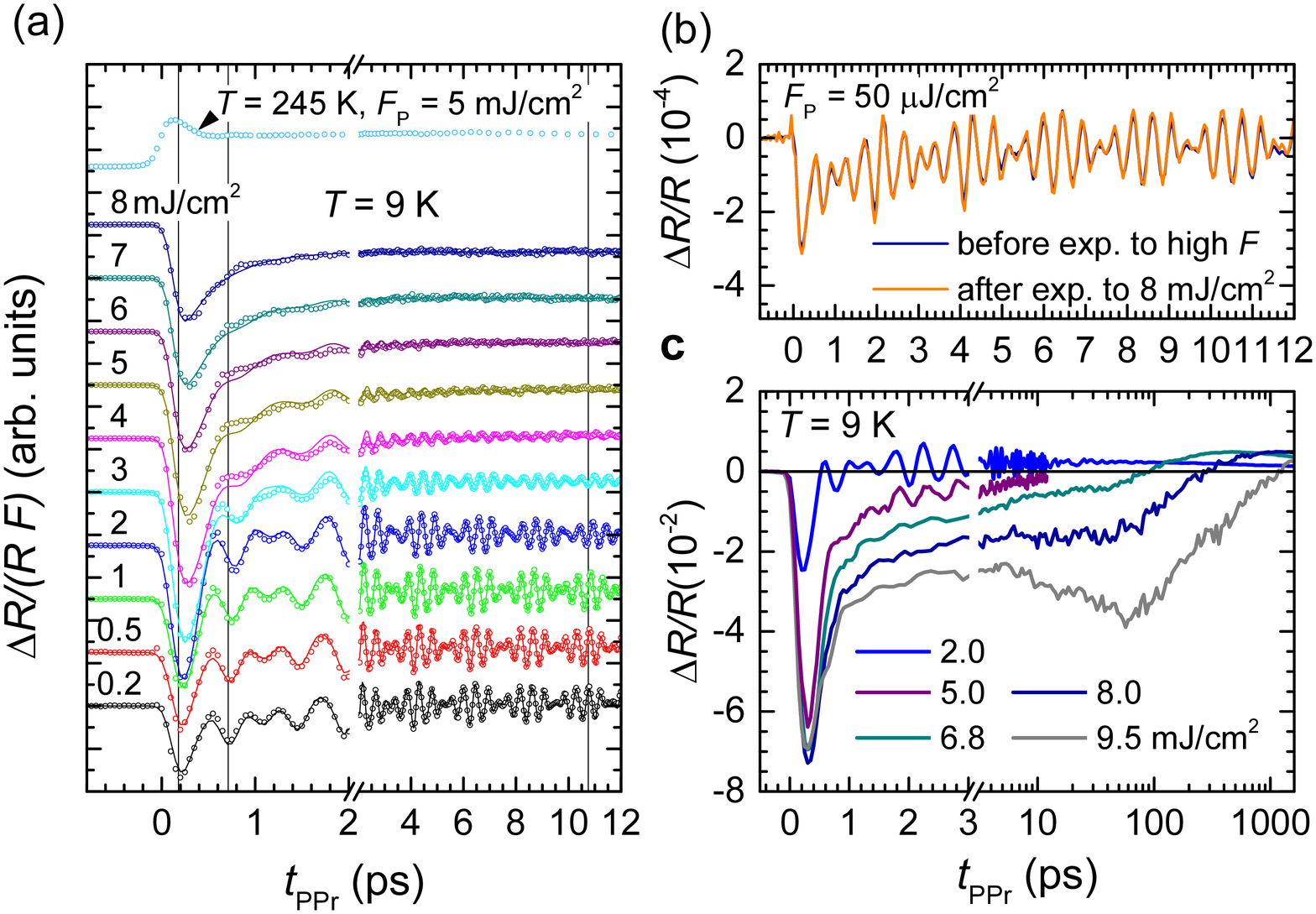}\caption{Transient reflectivity pump-fluence dependence. (a) Fluence-normalized
transient reflectivity as a function of $F_{\mathrm{P}}$ in the low-$T$
phase. A high-$T$ MC-phase trace ($T=245$ K) is shown on top for
comparison. Due to the normalization the absence of changes at low
$F_{\mathrm{P}}$ indicates linear scaling with $F_{\mathrm{P}}$.
The traces are vertically offset for clarity and the thin lines correspond
to the DECP model fit. (b) The low-$F_{\mathrm{P}}$ transient reflectivity
before and after exposure at the highest $F_{\mathrm{P}}$. (c) Transient
reflectivity at elevated $F_{\mathrm{P}}$ at longer delay. Please
note the logarithmic scale after the break.}
\label{fig:DRvsF} 
\end{figure}

\begin{figure}
\includegraphics[clip,width=1\columnwidth]{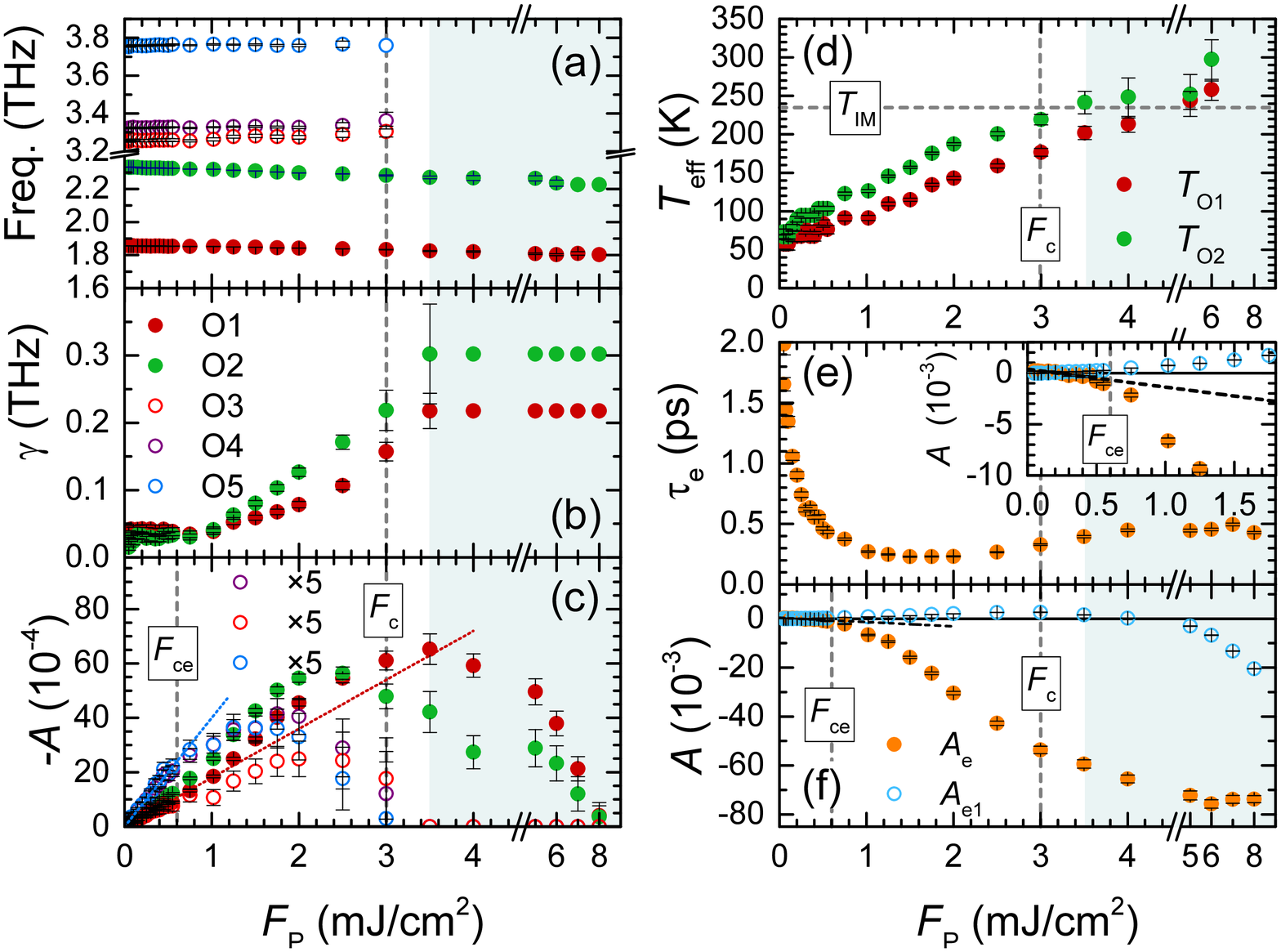}

\caption{Pump fluence dependence of the DECP fit parameters. (a)\textbf{ }The
frequency,\textbf{ }(b)\textbf{ }damping and (c)\textbf{ }amplitude
of the coherent oscillations as a function of $F_{\mathrm{P}}$. In
the shaded region, above $F_{\mathrm{P}}\sim3.5$~mJ/cm$^{2}$, the
coherent oscillations cannot be completely described by the DECP theory.
(d) The effective temperature (see text) obtained from the frequency
shifts of the two strongest coherent modes as a function of $F_{\mathrm{P}}$.
(e)\textbf{ }The\textbf{ }relaxation time and (f) amplitude of the
exponential components as a function of $F_{\mathrm{P}}$. The black
dashed line is the extrapolation of the low $F_{\mathrm{P}}$ behavior.
The inset to (e) shows (f) in expanded scale. The vertical and horizontal
dashed lines in correspond to $F_{\mathrm{ce}}$, $F_{\mathrm{c}}$
and $T_{\mathrm{IM}}$, respectively.\label{fig:FPvsF}}
\end{figure}

In Figure \ref{fig:DRvsF} (a) we show the pump fluence, $F_{\mathrm{P}}$,
dependence of the standard two-pulse transient reflectivity, $\Delta R/R$,
at $T=9$ K. The CO scale almost linearly with increasing $F_{\mathrm{P}}$
up to $\sim2$ mJ/cm$^{2}$. With increasing $F_{\mathrm{P}}$ the
CO appear suppressed in the $F_{\mathrm{P}}$-normalized scans and
the signal becomes dominated by the sub-ps relaxation component. The
detailed DECP-model\citep{NaseskaSutar2020,naseskaSutar2020suppl}
component analysis (see. Figure \ref{fig:FPvsF}) reveals that the
sub-ps exponential component (SEC) amplitude, $A\mathrm{_{e}}$, departs
from the low-$F_{\mathrm{P}}$ linear dependence and starts to increase
more steeply above $F_{\mathrm{ce}}\sim0.6$ mJ/cm$^{2}$ already
(see inset to Figure \ref{fig:FPvsF} (e)). Concurrently, the amplitudes
of the three weaker, high-frequency modes, designated O3, O4 and O5,
show an onset of saturation while the amplitudes of the two stronger,
low-frequency modes, O1 and O2, show an increased slope. Above $F_{\mathrm{c}}\sim3$
mJ/cm$^{2}$ the amplitudes of all modes drop and vanish at the highest
$F_{\mathrm{P}}$. The SEC does not show any anomalies around $F_{\mathrm{c}}$,
but shows saturation of $A_{\mathrm{e}}$ and the relaxation time,
$\tau_{\mathrm{e}}$, above $F\mathrm{_{P}\sim5}$ mJ/cm$^{2}$.

The high-$F$ transient reflectivity shows an initial fast, $\tau_{\mathrm{e}}=0.5$
ps, relaxation followed by slower $F$-dependent dynamics (see Figure
\ref{fig:DRvsF} (c)). The slow dynamics sets in above $F_{\mathrm{P}}\sim5$
mJ and consists of a plateau emerging into a broad dip at $\sim60$
ps at the highest $F_{\mathrm{P}}$ followed by relaxation extending
to a $\sim1$ ns timescale.

\begin{figure}
\includegraphics[clip,width=0.8\columnwidth]{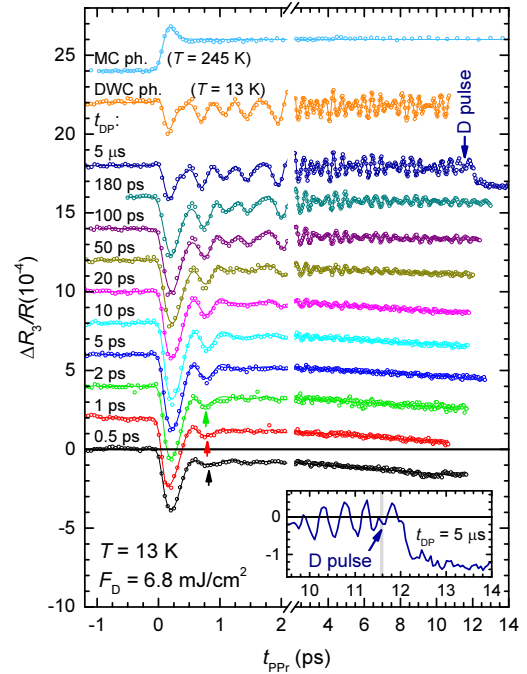}\caption{Multi-pulse transient reflectivity, $\Delta R_{3}/R$, as a function
of the delay between the D and P pulses, $t_{\mathrm{DP}}$. The arrows
at the short-$t_{\mathrm{DP}}$ scans emphasize the CO while the arrow
at the $t_{\mathrm{DP}}=5$-$\mu$s scan indicates the arrival of
the subsequent D pulse. The two top-most traces, shown for comparison,
correspond to the thermal equilibrium MC- and DWC-phase transient
reflectivity. The traces are vertically offset for clarity and the
thin lines in both panels correspond to the DECP model \citep{zeiger1992theory}
fit (see SM \citep{naseskaSutar2020suppl}).\label{fig:DrvstDP}}
\end{figure}

The high-$F_{\mathrm{P}}$ slow dynamics was investigated further
by means of the multi-pulse experiments. The multi-pulse transient
reflectivity, $\Delta R_{3}/R$, well above $F_{\mathrm{c}}$ is shown
in Figure \ref{fig:DrvstDP} for different D-P delays, $t_{\mathrm{DP}}$.
At short $t_{\mathrm{DP}}$, the multi-pulse transient reflectivity
resembles the 2-pulse transient reflectivity of the low-$T$ state
around $F_{\mathrm{P}}\sim F_{\mathrm{c}}$, but, with \emph{much
strongly damped} CO. After $t_{\mathrm{DP}}\sim5$ ps the CO start
to gradually reemerge also on longer P-Pr delays, $t_{\mathrm{PPr}}$.
The multipulse transients are \emph{completely different} from the
high-$T$ MC phase transients indicating that the highly excited transient
state cannot be associated with the high-$T$ MC phase.

The DECP fit component analysis shown in Figure \ref{fig:FPvstDP}
indicates that the initial amplitudes of the strongest two modes (see
Figure \ref{fig:FPvstDP} (c)) \emph{do not appear suppressed} even
immediately after the D pulse while the damping factors (see Figure
\ref{fig:FPvstDP} (b)) exceed the equilibrium value \citep{NaseskaSutar2020}
of $\sim0.01$ THz by up to two orders of magnitude recovering by
an order of magnitude on a $\sim100$ ps $t_{\mathrm{DP}}$ timescale.
Just before the arrival of the subsequent D pulse, after 5 $\mu$s,
the damping is reduced close to the equilibrium-state value, together
with exponential components parameters (Figure \ref{fig:FPvstDP}
(e) and (f)). The low-$T$ state is therefore completely recovered
after each D pulse, albeit, at a higher $T$ than the cryostat base
temperature due to the heat buildup. The behavior is similar also
at an increased $F_{\mathrm{D}}=8.1$ mJ/cm$^{2}$ with slower recovery.
\citep{naseskaSutar2020suppl}

The suppression of the CO is delayed with respect to the D pulse for
$\sim0.6$ ps (see inset to Figure \ref{fig:DrvstDP}) and does not
depend on the small changes of the $t_{\mathrm{DP}}$ delay \citep{naseskaSutar2020suppl}.
The delay is consistent with the duration of the initial sub-picosecond
transient observed for the high-$F$ transient reflectivity in Figure
\ref{fig:DRvsF} (a).

No irreversibility of the transient reflectivity was observed after
exposing the sample to long trains of pump pulses at the highest $F_{\mathrm{P}}$
as shown in Figure \ref{fig:DRvsF}b. To check for any heat build
up effects single D-pulse experiments were also performed. No long-lived
changes of the ultrafast transient reflectivity were found.\citep{naseskaSutar2020suppl}

\section{Discussion}

The onset of the nonlinearities with increasing $F_{\mathrm{P}}$,
at $F_{\mathrm{ce}}\sim0.6$ mJ/cm$^{2}$, is mostly expressed in
the SEC that can be associated with the electronic degrees of freedom
only. The SEC total duration (including the risetime) of $\sim500$
fs significantly exceeds the time resolution of the setup ($\sim80$
fs) so the nonlinearity cannot be associated with the pump optical-transition
nonlinearity, but is associated with the low-energy electronic degrees
of freedom.

\begin{figure}
\includegraphics[clip,width=1\columnwidth]{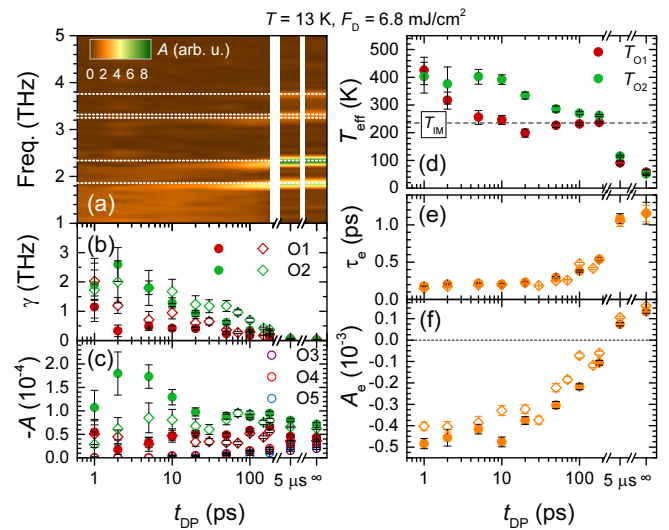}\caption{D-P delay dependence of the DECP fit parameters. (a) A density plot
of the Fourier-transform spectra from data in Figure \ref{fig:DrvstDP}.
The horizontal lines correspond to the equilibrium CO frequencies.
(b)\textbf{ }The CO dampings and (c) the corresponding amplitudes
as a function of $t_{\mathrm{DP}}$. (d) The effective temperature
obtained from the frequency shifts of the two strongest coherent modes
as a function of $t_{\mathrm{DP}}$. (e) The\textbf{ }relaxation time
and (f) amplitude of the exponential components as a function of $t_{\mathrm{DP}}$.
Open\textbf{ }diamonds in (b), (c), (e) and (f) correspond to the
experiment at $F_{\mathrm{D}}=8.1$ mJ/cm$^{2}$.\label{fig:FPvstDP}}
\end{figure}

Focusing on the lattice degrees of freedom, which show strong $F_{\mathrm{P}}$
nonlinearity on the larger fluence scale above $F_{\mathrm{c}}$,
we first estimate the transient heating taking the $T$-dependent
frequencies of the strongest two coherent modes,\citep{NaseskaSutar2020}
O1 and O2 at 1.86 THz and 2.34 THz, respectively, as proxies for the
lattice temperature. The obtained effective temperatures (see Figure
\ref{fig:FPvsF} (d)) reach $T_{\mathrm{IM}}$ at a fluence that is
close to $F_{\mathrm{c}}$. This indicates that the symmetry-breaking
lattice distortion is significantly affected only after the lattice
degrees of freedom are heated close to $T_{\mathrm{IM}}$. This is
consistent with the absorbed energy at $F_{\mathrm{c}}$ that is comparable
to the equilibrium enthalpy difference\citep{naseskaSutar2020suppl}
between the high-$T$ MC phase at $T\sim T_{\mathrm{IM}}$ and the
low-$T$ phase at $T\sim100$ K. Here the initial state $T$ at $t_{\mathrm{DP}}\sim5$
$\mu$s (see Figure \ref{fig:FPvstDP} (d)) is also estimated from
the coherent mode frequencies.

The suppression of the CO with increasing $F$ above $F\mathrm{_{c}}\sim3$
mJ/cm$^{2}$ is not abrupt as in the equilibrium case. This can be
attributed to the finite-optical-penetration-depth induced excitation
inhomogeneity. The thickens of a thin surface layer, within which
the absorbed energy threshold is exceeded, reaches the optical penetration
depth thickness only at the external fluence of $e\times F_{\mathrm{c}}\sim8$~mJ/cm$^{2}$,
consistent with the fluence where the CO are completely suppressed
(see Figure \ref{fig:FPvsF} (c)). The excitation inhomogeneity results
in a rather large error bar of $F\mathrm{_{c}}$, of the order of
$\pm0.5$ mJ/cm$^{2}$, but does not hinder a clear separation between
$F\mathrm{_{ce}}$ and $F\mathrm{_{c}}$.

The electronic ordering is therefore affected at lower excitation
densities than the lattice order. The drop of $\tau_{\mathrm{e}}$
from the equilibrium value of a few picoseconds to $\sim200$ fs indicates
that the gap-induced relaxation bottleneck \citep{NaseskaSutar2020}
is transiently suppressed above $F_{\mathrm{ce}}$ already. The estimated
photoexcited carriers density at $F_{\mathrm{ce}}$ is of the order\citep{naseskaSutar2020suppl}
of $10^{20}$ cm$^{-3}$ resulting in the plasma frequency of $\sim0.4$
eV that is larger than the insulating gap\citep{wangCao2004optical}
of $\Delta_{\mathrm{I}}\sim0.15$ eV. The bottleneck suppression can
therefore be attributed to a sub-picosecond transient washout of the
gap due to the (screening-induced) Mott transition that lasts at least
few hundred fs. On this timescale a significant amount of the photoexcited
carrier energy can be incoherently damped to phonons since the inelastic
electron-phonon scattering time can be as fast as $\sim10$ fs\citep{lundstrom_2000}.
However, the low frequency phonons corresponding to Ir-ions displacements
\citep{NaseskaSutar2020} that set the lattice order parameter dynamics
appear coherent until the gap is restored. According to Ref. {[}\citealp{khomskiiMizokawa2005}{]}
the equilibrium IM transition is a combination of orbital ordering
and Peierls charge density wave transition. The Ir-t$_{2g}$-orbitals
derived bands are split (band Jahn-Teller effect) into two fully occupied
$xz$- and $yz$-derived bands and a broader 3/4 occupied $xy$-derived
band that simultaneously becomes gaped due to the Peierls tetramerization
(see Fig \ref{fig:struct} (c)). While the photoexcited Mott transition
transiently suppresses the $xy$-derived band charge density wave,
the band Jahn-Teller effect remains effective stabilizing the broken
symmetry lattice deformation until the gap is restored. A similar
decoupling of the electronic and lattice orders on a short timescale
has been previously observed in TiSe$_{2}$. \citep{porerLeierseder2014}

Turning to the highly excited transient state (HETS), the multi-pulse
experiments clearly confirm that even at $F_{\mathrm{D}}\sim8$ mJ/cm$^{2}$
the excited volume does not switch into the high-$T$ MC state despite
the large absorbed energy. In the most excited part of the volume,
near the surface, the absorbed energy exceeds the equilibrium enthalpy
difference between the high-$T$ MC state just above $T_{\mathrm{IM}}$
and the initial low-$T$ state by \emph{more than two times}. The
presence of the strongly damped broken-symmetry \citep{NaseskaSutar2020}
CO with the unsuppressed initial amplitude (see Figure \ref{fig:FPvstDP}
(c)) indicates that the local lattice symmetry remains broken in most
of the excited volume, albeit with strong dephasing due to fluctuations
and/or increased disorder.

The broken symmetry state could be stabilized by the pressure exerted
by the surrounding unexcited bulk low-$T$ phase. Application of pressure
is known to stabilize the triclinic insulating phase \citep{oomiKagayama1995effect},
which has $\sim0.7$ \% larger density \citep{furubayashiMatsumoto1994}
than the MC phase. Taking the compressibility\citep{garg2007reentrant},
$\beta\sim6\cdot10^{-3}$ GPa$^{-1}$, one obtains the pressure of
$\sim1.1$ GPa leading to an increase\citep{oomiKagayama1995effect}
of $T\mathrm{_{IM}}$ to $T\mathrm{_{IM}}'\sim260$K.

In addition to the $T\mathrm{_{IM}}$ increase, the pump optical penetration
depth could increase due to the optical-transition bleaching at high-$F$
resulting in saturation of the absorbed energy density.\citep{naseskaPogrebna2018}
Such saturation cannot be entirely excluded and the maximum experimental
$T_{\mathrm{eff}}$ indeed does not appear to significantly exceed
$T\mathrm{_{IM}}'$ (Figures \ref{fig:FPvsF} (d) and \ref{fig:FPvstDP}
(d)). We should note, however, that at the high excitation $T_{\mathrm{eff}}$
does not directly correspond to the most excited volume region since
the CO are inhomogeneously suppressed so $T_{\mathrm{eff}}$ is somewhat
weighted towards the less-excited volume region.

While the above effects can contribute to stabilization of the broken
symmetry state, they \emph{cannot account} for the observed large
CO dephasing. Extrapolating the equilibrium $\gamma(T)$\citep{NaseskaSutar2020}
to the estimated\citep{naseskaSutar2020suppl} maximum possible transient
$T\sim400$ K we obtain $\gamma\sim0.08$ and $\sim0.16$ for the
O1 and O2 mode, respectively, which are at least$\sim5$ times smaller
than the observed $\gamma$ below $t_{\mathrm{DP}}\sim20$ ps (Figure
\ref{fig:FPvstDP} (b)). The picosecond transient state therefore
cannot be characterized as an ordinary super-heated equilibrium pressure-stabilized
low-$T$ phase.

According to the classical kinetic theory of first order phase transitions
\citep{lifshitz1995physical} the formation of a stable phase from
a metastable one proceeds through formation of macroscopic droplets
that, due to the surface energy terms, grow deterministically only
if their size exceeds a critical size, $R_{\mathrm{c}}$. The nucleation
of the droplets with $R>R_{\mathrm{c}}$ is a stochastic process driven
either by fluctuations or tunneling, the later being proposed for
the broken-symmetry-state droplet formation in the supercooled early
universe \citep{linde1982new}. In the present case thermal fluctuations
are likely large enough that the tunneling should not play any role.

The observed HETS could, therefore, be tentatively understood in this
context where the fluctuating droplets of the high-$T$ MC phase introduce
strong dynamical disorder into the triclinic phase matrix, but fail
to reach the critical size or their growth is too slow to reach the
MC phase before the heat diffusion spreads the absorbed energy across
a larger volume, recovering the stability of the triclinic phase.
Since there was no observable variation of the pump scattering with
$t_{\mathrm{DP}}$ in the multipulse experiments the droplets are
presumably much smaller than the pump wavelength of 800 nm and are
most likely size-limited by the characteristic low-$T$-phase twin
lamella thickness \citep{sunKimoto2001situ} of $\sim20$ nm.

Increasing the excitation fluence beyond $F\sim8$ mJ/cm$^{2}$ the
standard 2-pulse $\Delta R/R$ (see Fig. \ref{fig:DRvsF} (c)) shows
evolution of a broad dip between $\sim10$ and $\sim200$ ps. The
dip might signify an onset of MC-phase droplets growth. Unfortunately,
this excitation region was not investigated in multipulse experiments
in order to avoid sample cracking\footnote{Observed in earlier experiments upon high laser induced average thermal
load.} due to the high D-beam average thermal load\footnote{The average thermal load in the multipulse experiments is doubled
with respect to the standard 2-pulse experiments since D-beam is unchopped.}. Further experiments with lower repetition laser source are therefore
necessary to reach a more definitive conclusion regarding the behavior
beyond $F\sim8$ mJ/cm$^{2}$.

The HETS is expected to be quite metallic and different from the DWC
state since the sub-picosecond component behavior indicates the suppression
of the gap at lower $F$ already. However, the evidence is indirect
so more direct experimental evidence from time-resolved low-energy
probes is necessary to clearly establish the degree of metallicity.

Since the HETS state appears after strong ultrafast excitation on
a sub-picosecond timescale, in a manner similar to the metastable
metallic H phase \citep{stojchevskaVaskivskyi2014ultrafast,Ravnik2019}
in 1$T$-TaS$_{2}$, it is necessary to compare the phenomena. 1$T$-TaS$_{2}$
is, unlike CuIr$_{2}$S$_{4}$, a layered compound with the equilibrium
phase diagram that is significantly more complicated than the phase
diagram of CuIr$_{2}$S$_{4}$. In particular, the low-$T$ insulating
commensurate (C) polaron ordered phase is separated from the symmetric
high-$T$ metallic phase ($T>543$ K) by intermediate incommensurate
polaron ordered phases that are quite conducting.

The H phase is created after the C phase exposure to a single ultrafast
optical pulse above a critical fluence, $F_{\mathrm{H}}\sim1$ mJ/cm$^{2}$.\citep{stojchevskaVaskivskyi2014ultrafast}
On the long timescales the H phase appears as a disordered texture
of the frozen C-phase polaron domains\citep{gerasimenkoKarpov2019intertwined}.
An additional metastable amorphous (A) polaron glass phase has also
been observed recently that is created at higher excitation fluences
than the H phase, $F\gtrsim F_{\mathrm{A}}\sim3.5$ mJ/cm$^{2}$.\citep{gerasimenkoVaskivskyi2019quantum}

The spatial textures of the H and A phases and their evolution on
the ultrafast timescales have not been yet determined and the microscopic
pathway from the C to either H or A phase is still unclear. Since
different C-phase domains in the H phase have different phase shifts
relative to the underlying atomic lattice\citep{gerasimenkoKarpov2019intertwined}
a global reconfiguration of the polaron lattice must take place during
the transition already above $F_{\mathrm{H}}$. This implies that
the C polaron lattice ``melts'' into some transient state before
condensing into the H (A) phase. This transient state could be, similarly
to the present case on the picoseconds timescales, composed from droplets
of a ``melted'' phase that are embedded into the C phase. The droplet
phase could correspond to either some incommensurate phase, polaron
liquid or even the high-$T$ metallic phase without polarons.

The CO in 1$T$-TaS$_{2}$ are not so strongly suppressed during the
creation of the H phase and become suppressed only for $F>F_{\mathrm{A}}.$\citep{ravnikVaskivskyi2018real,ravnikDiego2020time}
The H phase creation conditions in 1$T$-TaS$_{2}$ are therefore
more similar to the excitation above $F_{\mathrm{ce}}$ in CuIr$_{2}$S$_{4}$,
where the lattice $T$ remains below $T\mathrm{_{IM}}$, while the
fluences above $F_{\mathrm{c}}$, where also the lattice $T$ transiently
exceeds $T\mathrm{_{IM}}$, correspond better to the A-phase creation
conditions in 1$T$-TaS$_{2}$. Owing to the similar penetration depths
of $\sim30$ nm and $\sim40$ nm in 1$T$-TaS$_{2}$\citep{gerasimenkoVaskivskyi2019quantum}
and CuIr$_{2}$S$_{4}$\citep{naseskaSutar2020suppl}, respectively,
and the similar equilibrium transition temperatures it is not surprising
that $F_{\mathrm{c}}$ and $F_{\mathrm{A}}$ are rather similar.

Contrary to 1$T$-TaS$_{2}$ in CuIr$_{2}$S$_{4}$ the same DWC state
forms on the long timescales irrespective of the excitation fluence
and the excitation type. It was shown\citep{kiryukhinHoribe2006,bozinMasadeh2011}
that in the DWC state the long range low-$T$ order is replaced by
a short range ($\xi\sim2$ nm) incommensurate modulation with preserved
\emph{local} Ir$^{4+}$ dimerization. The DWC state is, therefore,
structurally more comparable to the A phase than the H phase of 1$T$-TaS$_{2}$.
However, the electronic gap in the DWC state is not suppressed\citep{takuboMizokawa2008}
like in the A state of 1$T$-TaS$_{2}$\citep{gerasimenkoVaskivskyi2019quantum}
and the resistivity is larger, $\rho\sim65$ $\Omega$cm, than in
the A state of 1$T$-TaS$_{2}$ with $\rho$ in the $\sim10$ m$\Omega$
range\citep{gerasimenkoVaskivskyi2019quantum}. The DWC and A state
(as well as H state) are therefore quite different. This could be
connected to the absence of strong correlations and weak neighboring
Ir-ion chain coupling in CuIr$_{2}$S$_{4}$ making the electronic
gap less sensitive to the disorder-induced doping.

Comparing CuIr$_{2}$S$_{4}$ with VO$_{2}$ and V$_{2}$O$_{3}$
we observe similarities, but also differences. All three compounds
show transient intermediate states with suppressed gaps \citep{morrisonChatelain2014photoinduced,wegkampHerzog2014instantaneous,lantzMansart2017ultrafast,singerRamirez2018nonequilibrium}
that are structurally distinct from the high-$T$ metallic phases
suggesting a transient decoupling of the structural and electronic
orders. The decoupling in both oxides has been attributed to the correlation
effects \citep{wegkampHerzog2014instantaneous,lantzMansart2017ultrafast}.
In our case, however, the effect can be attributed to a photoinduced
Mott transition.

The timescales of the structural transitions are quite different.
In VO$_{2}$ the transition is abrupt \citep{jager2017tracking,WallYang2019,OttoCotret2019}
while V$_{2}$O$_{3}$ \citep{singerRamirez2018nonequilibrium,ronchiHomm2019early}
and CuIr$_{2}$S$_{4}$ show first-order kinetics bottlenecks which
slow down the transition. In this context it is worth noting that
the lattice volume change at the equilibrium IM transition is the
smallest, 0.1 \% in VO$_{2}$ \citep{marezioDernier1970x} in comparison
to -1.3 \% in V$_{2}$O$_{3}$ \citep{rozierRatuszna2002comparative}
and 0.7~\% in CuIr$_{2}$S$_{4}$ \citep{furubayashiMatsumoto1994},
suggesting a smaller inter-phase-boundary energy cost enabling easier
nucleation of the high-$T$ metallic phase droplets.

\section{Conclusions}

We showed that in the strongly nonequilibrium regime upon a femtosecond
photoexcitation the low-temperature ultrafast structural dynamics
in CuIr$_{2}$S$_{4}$ is dominated by the first-order-transition
nucleation kinetics that prevents the complete ultrafast photoinduced
structural transition into the high-$T$ phase at unexpectedly large
excitation densities. Concurrently, the dynamically-decoupled electronic
order is suppressed rather independently on a sub-picosecond timescale
at much weaker excitation density. While the electronic order suppression
likely results in a transient metallization of the highly nonequilibrium
broken lattice-symmetry state no evidence was found that femtosecond
excitation would lead to a long lived metastable state more conducting
than the slowly formed disordered weakly conducting phase.

\section*{Acknowledgments}

The authors acknowledge the financial support of Slovenian Research
Agency (research core funding No-P1-0040 and young researcher funding
No. 50504) for financial support. We would also like to thank V. Nasretdinova
and E. Goreshnik for the help at the sample characterization, M. Ani\v{c}in
and A. Bavec for help at transient reflectivity measurements and E.
Bozin for fruitful discussions.

\bibliographystyle{apsrev4-1}
\bibliography{biblio}

\end{document}